\documentclass{aa}
\usepackage{graphicx}

\begin{document}

   \title{Detection of HD in emission towards \object{Sagittarius~B2}
          \thanks{Based on observations with ISO, an ESA project with instruments funded by ESA Member States (especially  the PI countries: France, Germany, the Netherlands and the United Kingdom) with the participation of ISAS and NASA.}
   }

   \author{E. T. Polehampton
           \inst{1}\fnmsep\inst{2}
           \and
           J.-P. Baluteau
	   \inst{3}
           \and
	   C. Ceccarelli
	   \inst{4}\fnmsep\inst{5}
           \and
	   B. M. Swinyard
	   \inst{2}
           \and
	   E. Caux
           \inst{6}
           }

   \offprints{E. Polehampton:  \email{etp@astro.ox.ac.uk}}

   \institute{Astrophysics, University of Oxford, Keble Road, Oxford, OX1 3RH, UK
   \and 
     Rutherford Appleton Laboratory, Chilton, Didcot, Oxfordshire, OX11 0QX, UK
   \and
     Laboratoire d'Astrophysique de Marseille, CNRS \& Universit\'e de Provence, BP 8, F-13376 Marseille Cedex 12, France
   \and
     Observatoire de Bordeaux, BP 89, F-33270 Floirac, France
   \and
     Laboratoire d'Astrophysique, Observatoire de Grenoble, BP 53, F-38041 Grenoble Cedex 09, France
   \and
     CESR CNRS-UPS, BP 4346, F-31028 Toulouse Cedex 04, France
      }

   \date{Received  / accepted }

 \abstract{
The giant molecular cloud Sagittarius~B2, located near the Galactic Centre, has been observed in the far-infrared by the ISO Long Wavelength Spectrometer. Wavelengths in the range 47--196~$\mu$m were covered with the high resolution Fabry-P\'{e}rot spectrometer, giving a spectral resolution of 30--40~km~s$^{-1}$. The $J=1\rightarrow0$ and $J=2\rightarrow1$ rotational transitions of HD fall within this range at 112~$\mu$m and 56~$\mu$m. A probable detection was made of the ground state $J=1\rightarrow0$ line in emission but the $J=2\rightarrow1$ line was not detected above the noise. This allowed us to calculate an upper limit on the temperature in the emitting region of approximately 80~K and a value for the deuterium abundance in the \object{Sgr~B2} envelope of D/H=(0.2--11)$\times10^{-6}$. 

                 }

   \titlerunning{Detection of HD in emission towards \object{Sgr~B2}}
   \maketitle
%
%________________________________________________________________

\section{Introduction}

The abundance of deuterium in the interstellar medium is a fundamental question in astronomy.  Deuterium was only produced in significant amounts in the big bang and is destroyed via processing in stellar interiors. Observations of the D/H ratio in the ISM provide lower limits to the primordial deuterium abundance. 

The Galactic Centre is the region of our galaxy with the largest star formation rate and contains the most highly processed material. Recent models of galactic chemical evolution (e.g. Matteucci et al. \cite{matteucci}) predict that the abundance of deuterium should be drastically reduced in the Galactic Centre due to the large stellar processing rates. However, recent observations of the deuterated molecules DCO$^+$ and DCN by Jacq et al. (\cite{jacq}) and Lubowich et al. (\cite{lubowich}) towards Galactic Centre molecular clouds show that deuterium is clearly present and not as depleted as the models suggest. The final determination of the exact amount of deuterium from these observations depends both on the line modelling and on the deuteration process for HCO$^{+}$ and HCN. Deuterium in these molecules is expected to be enhanced over the actual D/H ratio by fractionation effects which increase for lower temperatures. 

For this reason, a very good way to measure the deuterium abundance is by observing the HD ground state rotational transition at 112~$\mu$m. Observation of this line measures the HD column density and comparison with the relevant H$_2$ column density leads to a straightforward determination of the D/H ratio. 

We report on observations of the 112~$\mu$m HD line towards \object{Sagittarius~B2} (\object{Sgr~B2}), a giant molecular cloud complex located $\sim$100~pc from the Galactic Centre. The complex consists of several clusters of compact \ion{H}{ii} regions and dense molecular cores surrounded by a diffuse envelope. Its far-infrared spectrum is dominated by continuum emission from dust in the central clusters, with a colour temperature of about 35~K and a blackbody peak near 90~$\mu$m (eg. see the recent review by Ceccarelli et al. \cite{ceccarelli}).

\section{Observations and Data Reduction}

Sgr B2 was observed as part of a wide spectral survey using the ISO Long Wavelength Spectrometer (LWS) Fabry-P\'{e}rot (FP) mode (Clegg et al. \cite{clegg}). The whole LWS spectral range, from 47 to 196~$\mu$m, was covered using 36 separate observations with a spectral resolution of 30--40~km~s$^{-1}$. In each observation both the LWS FP and grating were scanned to cover a wide range in wavelength. The first description of the survey and its results on ammonia lines are reported in Ceccarelli et al. (\cite{ceccarelli}).

The LWS beam had an effective diameter of 78$\arcsec$ at 112~$\mu$m (Gry et al. \cite{gry}). The beam was centred at coordinates $\alpha=17^{\mathrm{h}}47^{\mathrm{m}}21.75^{\mathrm{s}}$, $\delta=-28\degr 23\arcmin 14.1\arcsec$ (J2000). This gave the beam centre an offset of 21.5$\arcsec$ from the main far-infrared peak, which occurs near the radio and mm source, \object{Sgr~B2}~(M) (Goldsmith et al. \cite{goldsmithb}). This pointing was used to exclude the source \object{Sgr~B2}~(N) from the beam.

During each observation the LWS FP and grating settings were optimised for the detector whose band pass filter included the wavelength range of interest. This was designated as the `prime' detector. However, all ten LWS detectors recorded data simultaneously in their own spectral ranges. The other nine detectors are known as `non-prime' and often recorded useful data that can complement the prime data. 

The region around the HD $J=1\rightarrow0$ rotational transition at a rest wavelength of 112.0725~$\mu$m (Evenson et al. \cite{evenson}) was covered in two separate observations using the LWS long wavelength FP (FPL) and detector LW2. The first was a prime observation made on 1997 April 5 (ISO revolution 506) and the second was a non-prime observation made on  1997 April 8 (ISO revolution 509). The non-prime observation had limited wavelength coverage, starting at 112.064~$\mu$m (velocity of $-23$~km~s$^{-1}$). The sampling interval used in both observations was a quarter of a spectral resolution element with each data point repeated three times. The resolution element, determined by interpolating between measurements made on the ground and in orbit, was found to be $\approx$33~km~s$^{-1}$.

Each observation was processed to the Standard Processed Data (SPD) level using the LWS pipeline version 8. This produced data still in engineering units (FP gap voltage and detector photocurrent). Further calibration was carried out interactively using routines from the LWS Interactive Analysis (LIA) package (Gry et al. \cite{gry}). The standard LIA tool for processing FP data, `FP\_PROC', was modified using improved algorithms and calibration files in order to calibrate the data in flux (these appeared as part of LIA version 10). Due to a large straylight contribution (caused by the strength of the source), the dark signal was determined by an algorithm that examined instances during on-source observations when the grating and FP were aligned to cut out transmission in the main beam. The final dark value is $\sim$3\% of the continuum photocurrent and so is not a significant source of error in the calibrated flux. The relative grating response in FP mode was recovered from the non-prime observations in the dataset and a correction was applied for contamination from adjacent FP spectral orders (for more details of these corrections, see Polehampton et al. \cite{polehampton}; Gry et al. \cite{gry}. Further information will appear in Baluteau et al. in preparation).

The two observations were carefully checked for glitches and transient affected scans and these were removed using the ISO Spectral Analysis Package (ISAP). The observations were matched in wavelength by correcting each one to the local standard of rest. The final wavelength alignment was determined by examining the position of the H$_{2}$O 113.540~$\mu$m line which occurred in both observations. The remaining uncertainty is 0.004~$\mu$m (or 11~km~s$^{-1}$) which corresponds to the error in absolute wavelength calibration (Gry et al. \cite{gry}). A small scaling factor (10\%) was applied to the flux in the non-prime observation to bring the continuum level into agreement with the prime observation. The two observations were then co-added. 

The largest errors in the flux at 112~$\mu$m were due to the extended nature of the source and its offset from the LWS optical axis. These can cause a large scale fracturing and bending of the spectrum on each detector (Lloyd \cite{lloyd}). A correction to account for these effects was not applied. To give an idea of the error on the flux value, the data were compared with a medium resolution LWS grating mode observation of \object{Sgr~B2} taken with the same pointing. After correction for saturation in the long wavelength detectors, this shows a flux level at 112~$\mu$m that is $\sim$30\% smaller than the FP mode flux level and this reflects the magnitude of the error in absolute flux. Other calibration effects in FP mode for which the errors can be quantified contribute a systematic error of $<$5\%. 

The region around the HD $J=2\rightarrow1$ rotational transition at a rest wavelength of 56.230~$\mu$m (Ulivi et al. \cite{ulivi}) was also observed as part of the spectral survey. This line was covered in three non-prime observations using FPL with detector SW2. These observations were processed and co-added in a similar way to the 112~$\mu$m observations. The systematic error in absolute flux is dominated by uncertainty in the dark signal which makes up a large fraction (40\%) of the continuum photocurrent. The dark could not be determined in the same way as at 112~$\mu$m and so an estimate was made by comparison with FP prime observations and the medium resolution LWS grating mode observation. The final error in flux is estimated to be not more than 30\%.

\section{Results}

\begin{figure}
\resizebox{\hsize}{!}{\includegraphics{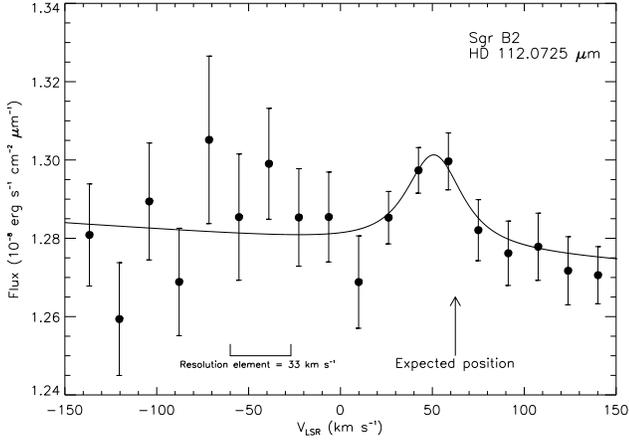}}
\caption{LWS high resolution Fabry-P\'{e}rot data in the region of the HD 112~$\mu$m line, plotted  against local standard of rest (LSR) velocity. The data have been binned at half a spectral resolution element. The error in each bin due to the spread in the data points is shown. Systematic errors in flux are not shown and are equal to roughly 30\%. The best fit to the line is shown after taking account of different offsets in the binning grid. The baseline has not been corrected for high frequency fringing (see text). The expected position of a line from \object{Sgr~B2} at +60--65~km~s$^{-1}$ is indicated.}
\label{HD_fit}
\end{figure}

Figure~\ref{HD_fit} shows the 112~$\mu$m data binned at half a spectral resolution element. Statistical error bars are shown for each bin and these are much smaller than the overall systematic error in the flux (30\%). The signal to noise clearly increases at velocities above $+25$~km~s$^{-1}$ due to the combined effect of overlap between FP mini-scans and the co-addition of non-prime data. A probable emission line detection occurs near the expected velocity of \object{Sgr~B2}~(M) (+60--65~km~s$^{-1}$; eg. Mehringer et al. \cite{mehringer}). In order to determine if this feature was real, individual mini-scans were examined from both observations as edge effects on FP scans can sometimes cause spurious features in the final spectrum. However, emission was present on all of the overlapping mini-scans before co-addition, showing that it is not due to scan edges. The co-added spectrum was checked for the high frequency fringing that is sometimes seen in LWS FP data (see Gry et al. \cite{gry}). The standard LIA routine, `Defringe', was used to determine the level of fringing, which could not account for the detected feature (this routine solves for the period, phase and amplitude of a sine wave fitted to the data). The fringing was not removed as it only affected the height of the detection by 5\%. The detection cannot be due leakage from adjacent FP spectral orders (see Gry et al. \cite{gry}) as there are no strong emission lines present in the nearby continuum. The 112~$\mu$m line has previously been detected by the LWS both in emission towards warm PDR regions in the \object{Orion Bar} (Wright et al. \cite{wright}) and in absorption towards the massive molecular cloud \object{W49N} (Caux et al. \cite{caux}). 

The line width of the absorption component due to \object{Sgr~B2} in the 6~cm transition of H$_{2}$CO has been observed to be $\approx$12~km~s$^{-1}$ (Mehringer et al. \cite{mehringerb}), which is unresolved by the LWS. Convolution of a Gaussian profile of this width and the LWS resolution element at 112~$\mu$m gives an expected line width $\approx$37~km~s$^{-1}$. Figure~1 shows the best fit of this convolved profile to the data. The effect of changing the position of the bin grid over the data was investigated and this did not change the fit parameters significantly. The best fit line is centred at a velocity of 51~km~s$^{-1}$ (consistent with \object{Sgr~B2} within the LWS FP wavelength calibration accuracy), with a line flux of $(3.3\pm1.4)\times10^{-12}$~erg~s$^{-1}$~cm$^{-2}$ (the error is 1$\sigma$ statistical uncertainty in the fit). Systematic uncertainty contributes a further error of $\sim$30\%.

The HD $J=2\rightarrow1$ transition at 56~$\mu$m was not detected in the co-added spectrum. An upper limit for emission in this line shows that the line flux must be less than 1.7$\times10^{-12}$~erg~s$^{-1}$~cm$^{-2}$.

\section{D/H Ratio}

Several studies on the dust and gas content of the diffuse cloud overlying \object{Sgr~B2}~(M) have been carried out. The structure of the regions around the source have recently been summarised by Ceccarelli et al. (\cite{ceccarelli}). Emission from HD could occur in two regions of the \object{Sgr~B2} complex which appear extended in the LWS beam. These are the warm (40--100~K) envelope, with $n$(H$_{2}$)$\approx 2\times10^{5}$~cm$^{-3}$, and the hot ($\sim$700~K) layer seen in NH$_{3}$ absorption with $n$(H$_{2}$)$\approx 10^{4}$~cm$^{-3}$.

At densities of $\sim$10$^{4}$~cm$^{-3}$, the deviation from LTE of the HD $J=1$ and $J=2$ level populations is very small (e.g. Bertoldi et al. \cite{bertoldi}). This means that the total HD column density can be derived from the line emission, $I_{\lambda}$, the source extent, $\Omega_\mathrm{source}$, and the gas temperature, $T_\mathrm{gas}$,
\begin{equation}
N(\mathrm{HD})=\frac{4\pi~I_{\lambda}}{h~\nu_{1,0}~A_{1,0}~\Omega_\mathrm{source}}~\frac{Q(T_\mathrm{gas})}{g_1\exp{(-E_{1}/k T_\mathrm{gas})}}
\end{equation}
where $E_1/k=128$~K, $A_{1,0}$ is the Einstein coefficient for spontaneous emission, equal to $5.12\times10^{-8}$~s$^{-1}$ (Abgrall, Roueff \& Viala \cite{abgrall}), $g_1$=3 and $Q(T_\mathrm{gas})$ is the partition function at the gas temperature. The angular size of the 100~$\mu$m continuum emission from the extended envelope is approximately $55\arcsec\times120\arcsec$ (Goldsmith et al. \cite{goldsmithb}), the centre of which was offset from the LWS optical axis by 21.5$\arcsec$. The solid angle covered is $\approx7\times 10^{-8}$~sr.

The observed flux at 112~$\mu$m can be used with the detection limit for the 56~$\mu$m line to constrain the temperature in the emitting region, assuming LTE. Comparing the population in the $J=1$ level with the upper limit for the population in the $J=2$ level leads to a temperature in the emitting region of $T_\mathrm{gas}<80$~K (taking account of errors). This excludes the possibility that the observed HD emission comes from the hot layer (it is unlikely that the hot region has a density much lower than 10$^4$~cm$^{-3}$, in which case the $J=2$ level would be subthermally populated).

Assuming, therefore, that the HD emission occurs in the warm envelope (with parameters from above, $n(\mathrm{H_{2}})\approx2\times10^{5}$~cm$^{-3}$ and $T$=40--80~K) gives a HD column density ranging from $1.8\times10^{18}$~cm$^{-2}$ (for $T$=80~K) to $6.0\times10^{18}$~cm$^{-2}$ ($T$=40~K). Taking account of 1$\sigma$ statistical and 30\% systematic error in the 112~$\mu$m line flux gives $N(\mathrm{HD})=(0.7$--$11)\times10^{18}$~cm$^{-2}$. 

Estimates of the molecular gas towards \object{Sgr~B2} have been obtained from studying absorption by trace molecules such as H$^{13}$CO$^+$ and H$^{13}$CN by Link et al. (\cite{link}). From these observations they derived $N$(H$_2$) between 4 and 5~$\times 10^{23}$~cm$^{-2}$. This estimate is consistent with studies of the dust extinction towards the central sources (Harvey et al. \cite{harvey}; Erickson et al. \cite{erickson}; Gatley et al. \cite{gatley}; Thronson \& Harper \cite{thronson}; Lis et al. \cite{lisa}) giving $N(\mathrm{H_2})=(0.5$--$2)\times10^{24}$~cm$^{-2}$ and observations of the whole \object{Sgr~B2} complex in sub-mm dust continuum (Goldsmith et al. \cite{goldsmitha}; Lis et al. \cite{lisb}; Gordon et al. \cite{gordon}; Kuan et al. \cite{kuan}). 

Adopting an H$_2$ column density of (0.5--2)$\times10^{24}$~cm$^{-2}$ gives a deuterium abundance of (0.2--11)$\times10^{-6}$, corresponding to temperatures in the emitting region of 80--40~K. These values may still be an underestimate, due to the possibility of self-absorption of HD within the \object{Sgr~B2} complex. Evaluating an accurate correction for self absorption would require a detailed knowledge of the various components in the \object{Sgr~B2} complex and is beyond the scope of this paper. However, approximate calculations show that absorption is unlikely to change the result by more than 50\%.

\section{Conclusions}

We report a probable emission line detection of the HD $J=1\rightarrow0$ transition at 112~$\mu$m towards \object{Sgr~B2}. An upper limit for the $J=2\rightarrow1$ transition at 56~$\mu$m is used to show that the emission must originate in the warm part of the \object{Sgr~B2} envelope, at a temperature lower than approximately 80~K. This leads to a deuterium abundance in the \object{Sgr~B2} complex of (0.2--11)$\times10^{-6}$. This range is constrained by the temperature (80--40~K) and H$_2$ column density ((0.5--2)$\times10^{24}$~cm$^{-2}$) in the emitting region. The error in D/H at a particular temperature and $N(\mathrm{H_2})$ within this range is approximately a factor of two.

The D/H ratio derived from DCO$^+$ and DCN observations of \object{Sgr~B2} by Jacq et al. (\cite{jacq}) was calculated to be in the range (0.35--10)$\times10^{-6}$, with a firm lower limit of 10$^{-7}$. Our values agree with this range and are consistent with the deuterium abundance of (1.7$\pm$0.3)$\times$10$^{-6}$, found in the \object{Sgr~A}~50~km~s$^{-1}$ cloud by Lubowich et al. (\cite{lubowich}).

The chemical evolution of the Galactic Bulge has been studied by Matteucci et al. (\cite{matteucci}). They predict that the present time D/H ratio in the bulge should be about three orders of magnitude smaller than in the Galactic Disk due to the shorter timescales for astration processes there. Our values show that D/H in \object{Sgr~B2} is possibly smaller than the average value found in the local interstellar cloud ((1.5$\pm$0.10)$\times$10$^{-5}$; Linsky \cite{linsky}) but cannot be as low as Matteucci et al. (\cite{matteucci}) predict. This may indicate that the deuterium content of the bulge has been enriched by infall of primordial material or that there could be a problem with the model. Prantzos (\cite{prantzos}) has calculated models that include infall and shows that the predicted D/H gradient is very sensitive to the assumed infall history.

The HD ground state rotational line provides a very reliable way to measure the deuterium abundance as $N$(HD) directly traces the deuterium column density. The remaining errors in the deuterium abundance are mostly due to uncertainties in the determination of the H$_{2}$ column density and temperature in the emitting region. Our probable detection of the 112~$\mu$m line indicates that with the higher sensitivity of SOFIA and maybe Herschel, many more observations of HD rotational lines will be possible. An improvement to the deuterium abundance derived from the HD line will also require an improvement in the determination of the H$_{2}$ column density. This could be carried out either by direct observation of the H$_{2}$ transitions in the near-infrared or by improving modelling of molecular abundances.

\begin{acknowledgements}

The ISO Spectral Analysis Package (ISAP) is a joint development by the LWS and SWS Instrument Teams and Data Centres. Contributing institutes are CESR, IAS, IPAC, MPE, RAL and SRON.  

\end{acknowledgements}


\begin{thebibliography}{}

\bibitem[1982]{abgrall}
Abgrall, H., Roueff, E., \& Viala, Y., 1982, A\&AS, 50, 505
\bibitem[1999]{bertoldi}
Bertoldi, F., Timmermann, R., Rosenthal, D., Drapatz, S., \& Wright, C. M., 1999, A\&A, 346, 267
\bibitem[2002]{caux}
Caux, E., Ceccarelli, C., Pagani, L., et al., 2002, A\&A, 383, L9
\bibitem[2002]{ceccarelli}
Ceccarelli, C., Baluteau, J.-P., Walmsley, M., et al., 2002, A\&A, 383, 603
\bibitem[1996]{clegg}
Clegg, P. E., Ade, P. A. R., Armand, C., et al., 1996, A\&A, 315, L38
\bibitem[1988]{evenson}
Evenson, K. M., Jennings, D. A., Brown, J. M., et al., 1988, ApJ, 330, L135
\bibitem[1977]{erickson}
Erickson, E. F., Caroff, L. J., Simpson, J. P., Strecker, D. W., \& Goorvitch, D., 1977, ApJ, 216, 404
\bibitem[1978]{gatley}
Gatley, I., Becklin, E. E., Werner, M. W., \& Harper, D. A., 1978, ApJ, 220, 822
\bibitem[1987]{goldsmitha}
Goldsmith, P. F., Snell, R. L., \& Lis, D. C., 1987, ApJ, 313, L5
\bibitem[1992]{goldsmithb}
Goldsmith, P. F., Lis, D. C., Lester, D. F., \& Harvey, P. M., 1992, ApJ, 389, 338
\bibitem[1993]{gordon}
Gordon, M. A., Berkermann, U., Mezger, P. G., et al., 1993, A\&A, 280, 208
\bibitem[2002]{gry}
Gry, C., Swinyard, B., Harwood, A., et al., 2002, ISO Handbook Volume IV (LWS), ESA SAI-99-077/Dc
\bibitem[1977]{harvey}
Harvey, P. M., Campbell, M. F., \& Hoffmann, W. F., 1977, ApJ, 211, 786
\bibitem[1999]{jacq}
Jacq, T., Baudry, A., Walmsley, C. M., \& Caselli, P., 1999, A\&A, 347, 957
\bibitem[1996]{kuan}
Kuan, Y., Mehringer, D. M., \& Snyder, L. E., 1996, ApJ, 459, 619
\bibitem[1981]{link}
Link, R. A., Stark, A. A., \& Frerking, M. A., 1981, ApJ, 243, 147
\bibitem[1998]{linsky}
Linsky, J. L., 1998, Space Sci. Rev., 84, 285
\bibitem[1991]{lisa}
Lis, D. C., Carlstrom, J. E., \& Keene, J., 1991, ApJ, 380, 429
\bibitem[1993]{lisb}
Lis, D. C., Goldsmith, P. F., Carlstrom, J. E., \& Scoville, N. Z., 1993, ApJ, 402, 238
\bibitem[2001]{lloyd}
Lloyd, C., 2001, in The Calibration Legacy of the ISO Mission, ESA SP-481, in press 
\bibitem[2000]{lubowich}
Lubowich, D. A., Pasachoff, J. M., Balonek, T. J., et al., 2000, Nature, 405, 1025
\bibitem[1999]{matteucci}
Matteucci, F., Romano, D., \& Molaro, P., 1999, A\&A, 341, 458
\bibitem[1993]{mehringer}
Mehringer, D. M., Palmer, P., Goss, W. M., \& Yusef-Zadeh, F., 1993, ApJ, 412, 684 
\bibitem[1995]{mehringerb}
Mehringer, D. M., Palmer, P., \& Goss, W. M., 1995, ApJS, 97, 497
\bibitem[2001]{polehampton}
Polehampton, E. T., Swinyard, B. M., Sidher, S. D., \& Baluteau, J.-P., 2001, in The Calibration Legacy of the ISO Mission, ESA SP-481, in press 
\bibitem[1996]{prantzos}
Prantzos, N., 1996, A\&A, 310, 106
\bibitem[1986]{thronson}
Thronson, H. A., \& Harper, D. A., 1986, ApJ 300, 396
\bibitem[1991]{ulivi}
Ulivi, L., De Natale, P., \& Inguscio, M., 1991, ApJ, 387, L29
\bibitem[1999]{wright}
Wright, C. M., van Dishoeck, E. F., Cox, P., Sidher, S. D., \& Kessler, M. F., 1999, ApJ, 515, L29

\end{thebibliography}
\end{document}